# Unexpectedly large mass loss during the thermal pulse cycle of the red giant R Sculptoris

Maercker, M.[1,2], Mohamed, S.[1,3], Vlemmings, W.H.T.[4], Ramstedt, S.[2], Groenewegen, M.A.T.[5], Humphreys, E.[1], Kerschbaum, F.[6], Lindqvist, M.[4], Olofsson, H.[4], Paladini, C.[6], Wittkowski, M.[1], de Gregorio-Monsalvo, I.[7], Nyman, L.-A.[7]

[1]European Southern Observatory, Karl-Schwarzschild-Str. 2, 85748 Garching, Germany
[2]Argelander Institute for Astronomy, University of Bonn, Auf dem Hügel 71, 53121 Bonn, Germany
[3]South African Astronomical Observatory, P.O. Box 9, Observatory 7935, South Africa
[4]Onsala Space Observatory, Dept. of Earth and Space Sciences, Chalmers University of Technology, SE-43992 Onsala, Sweden
[5]Royal Observatory of Belgium, Ringlaan 3, 1180 Brussels, Belgium
[6]University of Vienna, Dept. of Astrophysics, Türkenschanzstr. 17, 1180 Wien, Austria
[7]Joint ALMA Observatory, Chile

**The asymptotic giant branch star R Sculptoris is surrounded by a detached shell of dust and gas[1,2]. The shell originates from a thermal pulse during which the star undergoes a brief period of increased mass loss[3,4]. It has hitherto been impossible to constrain observationally the timescales and mass-loss properties during and after a thermal pulse – parameters that determine the lifetime on the asymptotic giant branch and the amount of elements returned by the star. Here we report observations of CO emission from the circumstellar envelope and shell around R Sculptoris with an angular resolution of 1.3". What was hitherto thought to be only a thin, spherical shell with a clumpy structure, is revealed to contain a spiral structure. Spiral structures associated with circumstellar envelopes have been seen previously, from which it was concluded that the systems must be binaries[5,6,7,8]. Using the data, combined with hydrodynamic simulations, we conclude that R Sculptoris is a binary system that underwent a thermal pulse ≈1800 years ago, lasting ≈200 years. About $3\times10^{-3}$ $M_\odot$ of mass was ejected at a velocity of 14.3 km s$^{-1}$ and at a rate ≈30 times higher than the pre-pulse mass-loss rate. This shows that ≈3 times more mass is returned to the interstellar medium during and immediately after a pulse than previously thought.**

The detached shell around R Sculptoris was observed in CO($J = 3 - 2$) emission at 345 GHz using the Atacama Large Millimeter/submillimeter Array (ALMA) during Cycle 0 operations (Fig.1, and supplementary information). The data clearly show the well-centered detached shell with a radius of 18.5", and reveal a spiral structure extending from the central star outwards to the shell. Previous observations of R Sculptoris show structure in the form of clumps. However, this was interpreted as clumpy material within the shell itself, and not as a structure interior to the shell[2].

Until now no clear signs of binary companions have been observed in the detached shell sources (with a possible exception for the detached shell around TT Cyg[9]). The observed structure around R Sculptoris, however, indicates the presence of a companion, shaping the stellar wind into a spiral shell structure[8]. Smoothed particle hydrodynamics (SPH) models show that a wide binary companion can have a significant effect in the shaping of the wind, leading to elliptical and spiral structures (e.g. as observed in the case of the envelope of AFGL 3068)[5,6].

The observed shapes of the circumstellar envelopes (CSEs) around binary AGB stars depend on the physical parameters of the binary system (e.g., separation and mass ratio[10]), the density contrasts imprinted on the wind, the temperatures in the CSE, the viewing angle, and, in the case of the gas, the chemistry and excitation[11]. The temporal variations of the mass-loss-rate and the expansion velocity further affect the structure of the CSE. Hence, the observed spiral structure and detached shell allow us to measure these important properties, and to directly link them to the thermal pulse.

Any change in the expansion velocity of the stellar wind will affect the spacing between the spiral windings. In Fig. 2 the spiral can be followed from the central star out to the detached shell over ≈5 windings. The 2.5 inner windings have a nearly constant spacing, with an average distance of 2.6", implying an essentially constant expansion velocity during the last 2.5 orbital periods. The expansion velocity of the present-day wind[3] of R Sculptoris gives an orbital period of $t_{orb}$ = 350 years. The position angle and radius of the observed emission then allow us to derive the velocity evolution of the stellar wind from R Sculptoris from the star out to the detached shell (Fig. 3, and supplementary information). The derived evolution of the expansion velocity since the last thermal pulse is consistent with models of thermal pulses[4]. However, the observed emission implies variations in the expansion velocity of ±1.5 km s$^{-1}$ on timescales of a few hundred years. Observed partial spiral windings and arcs, as well as emission at velocities up to 19 km s$^{-1}$, indicate brief periods of even larger velocity variations.

A spherically symmetric, detached shell can still be created in a binary system where the AGB star is undergoing a thermal pulse, due to the brief increase in mass-loss rate and expansion velocity. Collision with the surrounding, slower material will then shape the wind into a symmetric shell structure. The post-pulse mass loss leaves behind a spiral structure that connects the detached shell with the central star. Assuming a spherically symmetric expanding detached shell gives a shell expansion velocity of $v_{sh}$ = 14.3 km s$^{-1}$ and a shell radius of $R_{sh}$ = 18.5" (see supplementary information).

The present expansion velocity and size of the shell put the upper limit to the end of the thermal pulse to 1,800 years ago. With a binary period of 350 years we would expect to see ≈5 windings since the pulse, consistent with the observed spiral. A decelerated shell would imply a shorter time since the thermal pulse, and hence a shorter binary period or fewer spiral windings. Although a slight decrease in the expansion velocity of the detached shell is possible, we find no evidence of a *significant* decrease due to the sweeping-up of material from the pre-pulse wind. This is contrary to the current theory of how detached shells are formed during thermal pulses[3,4]. Also, no material is likely to have piled onto the shell due to the post-pulse mass loss.

Recent images of thermal dust emission show the detached shell, as well as a more distant, and spatially distinct, region of interaction with the interstellar material[12], showing the presence of a stellar wind from R Sculptoris before the pulse. Collision with a previous, slower wind is required in order to prevent the thin shell from quickly diffusing[4,13]. The average CO line intensity in the area surrounding the shell sets an upper limit to the pre-pulse mass-loss rate, and its ratio to the average CO line intensity in the shell suggests an increase in mass-loss rate during the thermal pulse of a factor of ≥10. The total shell-mass is essentially only due to the mass lost during the formation of the shell. An estimated total gas-mass in the shell of 2.5×10$^{-3}$ M$_\odot$ [3] gives a thermal pulse mass-loss rate of between 7×10$^{-6}$ M$_\odot$ yr$^{-1}$ and 2.5×10$^{-5}$ M$_\odot$ yr$^{-1}$ (see supplementary material), and implies a pre-pulse mass-loss-rate of ≤10$^{-6}$ M$_\odot$ yr$^{-1}$. The present-day mass-loss rate is estimated to be 3×10$^{-7}$ M$_\odot$ yr$^{-1}$,[3], i.e. a factor ≈30 lower than during the pulse. This general evolution of the mass-loss rate is consistent with stellar evolution models, however, the ratio between the derived pulse and pre-pulse mass-loss-rate is significantly higher than found in the models[4].

To further constrain the mass-loss-rate evolution of R Sculptoris, we modeled the system with a modified version of the GADGET-2 SPH code[14], including detailed radiative cooling[15]. The modeled system successfully forms a detached shell, including the observed spiral structure (see supplementary information for an animation). The modeled density, temperature, and velocity structures of the SPH model are then used as input in the 3-dimensional radiative transfer code LIME[16]. The global morphology of the modeled system closely resembles that of the observations, and the brightness distribution reproduces the observed intensities well (Fig. 4). We effectively constrain the mass-loss-rate evolution throughout the thermal pulse to the present time (Fig. 3). Assuming an inter-pulse time of 50,000 years (typical for stars of $\approx$1–4 $M_\odot$[17]) and our derived mass-loss-rate evolution, 10% of the mass lost between two subsequent pulses is expelled during the thermal pulse, and 40% during the first 1,800 years after the pulse. The pulse and immediate post-pulse phases are hence dominant in the formation and chemical enrichment of the CSE.

The chemical content of the expelled material depends critically on the physical properties of the pulses (e.g., pulse duration and interpulse mass-loss rate). The duration of the pulse limits the time for nucleosynthesis to occur inside the star[18], while the mass-loss rate between pulses limits the number of thermal pulses an AGB star will experience[17]. These properties will affect the stellar yields of new elements returned to the ISM, as well as eventually lead to the termination of the AGB phase. The observations presented here directly constrain these important physical parameters throughout the thermal-pulse cycle. In essence, it is the observed spiral structure that allows us to verify model results observationally, and refine our knowledge of thermal pulses and late stellar evolution.

**References**

[1]. Olofsson, H., Eriksson, K. & Gustafsson, B. SEST CO (J = 1 - 0) observations of carbon-rich circumstellar envelopes. *Astron. Astrophys.* **196**, L1–L4 (1988).

[2]. Olofsson, H., Maercker, M., Eriksson, K., Gustafsson, B. & Schöier, F. High-resolution HST/ACS images of detached shells around carbon stars. *Astron. Astrophys.* **515**, A27 (2010).

[3]. Schöier, F. L., Lindqvist, M. & Olofsson, H. Properties of detached shells around carbon stars. Evidence of interacting winds. *Astron. Astrophys.* **436**, 633–646 (2005).

[4]. Mattsson, L., Höfner, S. & Herwig, F. Mass loss evolution and the formation of detached shells around TP-AGB stars. *Astron. Astrophys.* **470**, 339–352 (2007).

[5]. Mastrodemos,N.&Morris,M.BipolarPre-PlanetaryNebulae:Hydrodynamicsof Dusty Winds in Binary Systems. II. Morphology of the Circumstellar Envelopes. *Astrophys. J.* **523**, 357–380 (1999).

[6]. Mauron, N. & Huggins, P. J. Imaging the circumstellar envelopes of AGB stars. *Astron. Astrophys.* **452**, 257–268 (2006).

[7]. Dinh-V.-Trung & Lim, J. Tracing the Asymmetry in the Envelope Around the Carbon Star CIT 6. *Astrophys. J.* **701**, 292–297 (2009).

[8]. Kim, H. & Taam, R. E. Probing Substellar Companions of Asymptotic Giant Branch Stars through Spirals and Arcs. *Astrophys. J.* **744**, 136 (2012).

[9]. Olofsson, H. *et al*. A high-resolution study of episodic mass loss from the carbon star TT Cygni. *Astron. Astrophys.* **353**, 583–597 (2000).

[10]. De Marco, O. The Origin and Shaping of Planetary Nebulae: Putting the Binary Hypothesis to the Test. *PASP* **121**, 316–342 (2009).

11. Politano, M. & Taam, R. E. The Incidence of Non-spherical Circumstellar Envelopes in Asymptotic Giant Branch Stars. *Astrophys. J.* **741**, 5 (2011).

12. Cox, N. L. J. *et al*. A far-infrared survey of bowshocks and detached shells around AGB stars and red supergiants. *Astron. Astrophys.* **537**, A35 (2012).

13. Steffen, M. & Schönberner, D. On the origin of thin detached gas shells around AGB stars. Insights from time-dependent hydrodynamical simulations. *Astron. Astrophys.* **357**, 180–196 (2000).

14. Springel, V. The cosmological simulation code GADGET-2. *Mon. Not. R. Astron. Soc.* **364**, 1105–1134 (2005).

15. Mohamed, S. & Podsiadlowski, P. Wind Roche-Lobe Overflow: a New Mass-Transfer Mode for Wide Binaries. In *15th European Workshop on White Dwarfs* (ed. R. Napiwotzki & M. R. Burleigh), vol. 372 of *Astronomical Society of the Pacific Conference Series*, 397–400 (2007).

16. Brinch, C. & Hogerheijde, M. R. LIME - a flexible, non-LTE line excitation and radiation transfer method for millimeter and far-infrared wavelengths. *Astron. Astrophys.* **523**, A25 (2010).

17. Karakas, A. & Lattanzio, J. C. Stellar Models and Yields of Asymptotic Giant Branch Stars. *PASA* **24**, 103–117 (2007).

18. Busso, M., Gallino, R. & Wasserburg, G. J. Nucleosynthesis in Asymptotic Giant Branch Stars: Relevance for Galactic Enrichment and Solar System Formation. *ARA&A* **37**, 239–309 (1999).

19. Olofsson, H., Carlstrom, U., Eriksson, K., Gustafsson, B. & Willson, L. A. Bright carbon stars with detached circumstellar envelopes - A natural consequence of helium shell flashes? *Astron. Astrophys.* **230**, L13–L16 (1990).

20. Olofsson, H., Eriksson, K., Gustafsson, B. & Carlstrom, U. A study of circumstellar envelopes around bright carbon stars. I - Structure, kinematics, and mass-loss rate. *ApJS* **87**, 267–304 (1993).

21. Claussen, M. J., Kleinmann, S. G., Joyce, R. R. & Jura, M. A flux-limited sample of Galactic carbon stars. *Astrophys. J. Supp.* **65**, 385–404 (1987).

22. Rosswog, S. Astrophysical smooth particle hydrodynamics. *NewAR* **53**, 78–104 (2009).

23. Springel, V. Smoothed Particle Hydrodynamics in Astrophysics. *ARA&A* **48**, 391–430 (2010).

24. Price, D. J. Smoothed particle hydrodynamics and magnetohydrodynamics. *Journal of Computational Physics* **231**, 759–794 (2012).

25. Barnes, J. & Hut, P. A hierarchical O(NlogN) force-calculation algorithm. *Nature* **324**, 446–449 (1986).

26. Smith, M. D. & Rosen, A. The instability of fast shocks in molecular clouds. *Mon. Not. R. Astron. Soc.* **339**, 133–147 (2003).

27. Mohamed, S., Mackey, J. & Langer, N. 3D simulations of Betelgeuse's bowshock. *Astron. Astrophys.* **541**, A1 (2012).

28. Ramstedt, S., Schöier, F. L., Olofsson, H. & Lundgren, A. A. On the reliability of mass-loss-rate estimates for AGB stars. *Astron. Astrophys.* **487**, 645–657 (2008).

29. De Beck, E. *et al*. Probing the mass-loss history of AGB and red supergiant stars from CO rotational line profiles. II. CO line survey of evolved stars: derivation of mass-loss rate formulae. *Astron. Astrophys.* **523**, A18 (2010).

**Supplementary Information**
Supplementary Information is linked to the online version of the paper at www.nature.com/nature.

**Acknowledgements**
This paper makes use of the following ALMA data:
ADS/JAO.ALMA#2011.0.00131.S. ALMA is a partnership of ESO (representing its member states), NSF (USA) and NINS (Japan), together with NRC (Canada) and NSC and ASIAA (Taiwan), in cooperation with the Republic of Chile. The Joint ALMA Observatory is operated by ESO, AUI/NRAO and NAOJ.
The authors thankfully acknowledge the technical expertise and assistance provided by the Spanish Supercomputing Network (Red Espanola de Supercomputacion), as well as the computer resources used: the LaPalma Supercomputer, located at the Instituto de Astrofisica de Canarias.
F.K. acknowledges funding by the Austrian Science Fund FWF under project number P23586-N16 and I163-N16, C.P. under P23006-N16.

**Author Contributions**
M.M planned the project, prepared and submitted the proposal, analyzed the data, and wrote the manuscript. S.M. was involved in project preparation, data interpretation, responsible for the SPH modeling, and commented on the manuscript. W.V. was involved in project planning, proposal preparation, data reduction and analysis, radiative transfer modeling, and commented on the manuscript. S.R. was involved in project planning, data analysis, and commented on the manuscript. The remaining authors were involved in the project preparation, science discussion, and commented on the manuscript.

**Author Information**
Reprints and permission information is available at www.nature.com/reprints

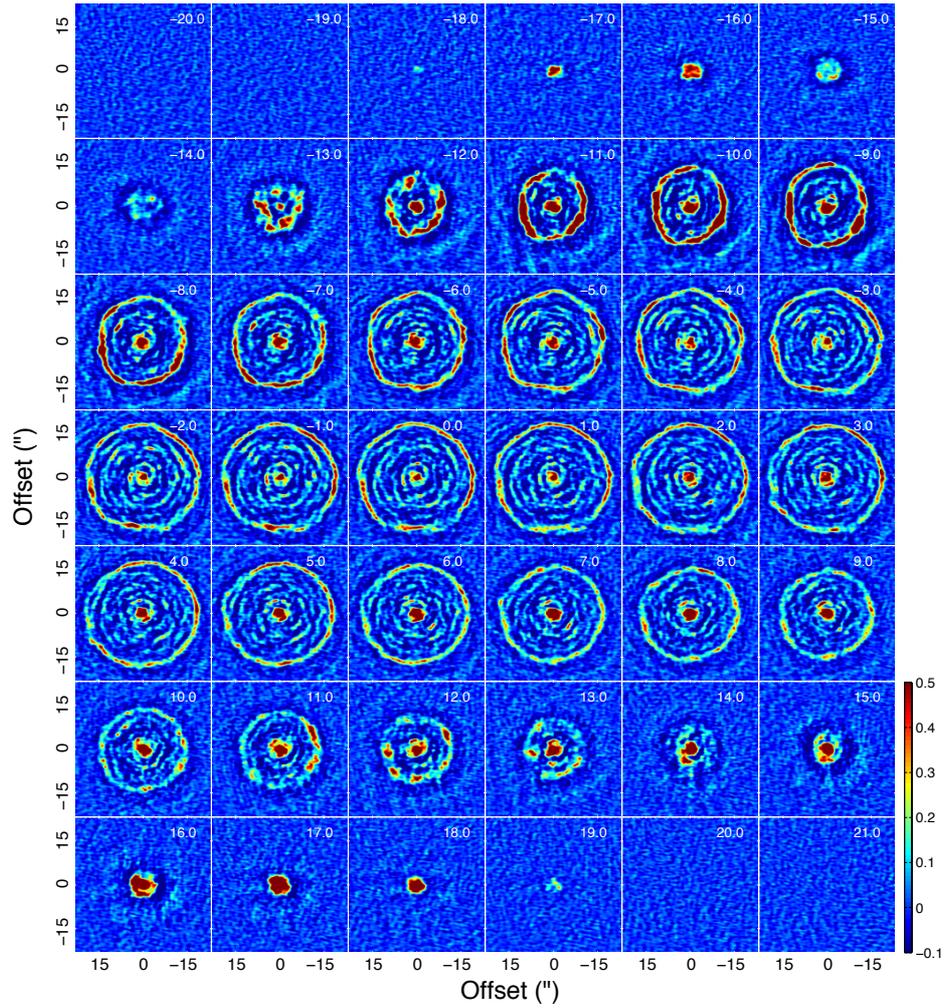

Figure 1: ALMA Early Science observations of the CO($J = 3 - 2$) emission from the AGB star R Sculptoris. The panels in the figure show the emission in the different velocity channels. The color scale is given in Jy/beam. The stellar velocity is at $v_{LSR}=-19$ km s$^{-1}$. The numbers in the top-right corners indicate the velocity in km s$^{-1}$ with respect to the stellar velocity. The spherical detached shell appears as a ring in the individual velocity channels, with its largest extent at the stellar velocity. The shell is clearly visible at 18.5" at the stellar $v_{LSR}$, as well as a spiral structure connecting the central star with the detached shell. The structure can be traced through all velocity channels.

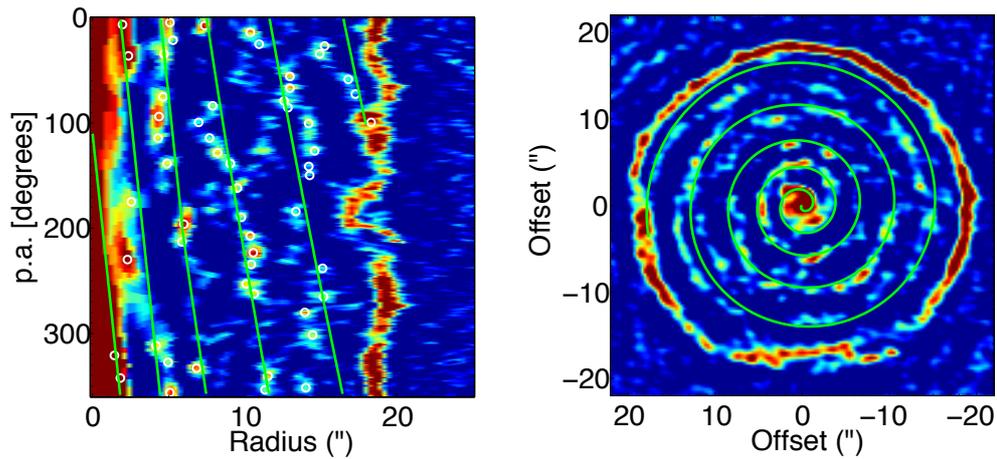

Figure 2: The CO($J = 3 - 2$) emission at the stellar $v_{LSR}$ of R Sculptoris. *Left:* Position angle (p.a.) vs. radius based on the stellar $v_{LSR}$ image. The p.a. starts in North and increases counter-clockwise. *Right:* The stellar $v_{LSR}$ image. The color scale is given in Jy/beam. The green lines show linear fits to the emission peaks (white circles) in the p.a. vs. radius diagram. The first 2.5 windings are nearly parallel, with a constant separation of 2.6"±0.07", indicating that the expansion velocity has been constant (on average) for the last 2.5 binary periods. The present-day expansion velocity is estimated to be 10.5 km s$^{-1}$, giving a binary period of 350 years. The linear fits can hence be translated directly into a velocity evolution (Fig. 3). The corresponding spiral is plotted on top of the stellar $v_{LSR}$ image (right). Deviations from a perfect spiral are on the order of ±1.5 km s$^{-1}$, indicating small velocity variations over times of ≈ 50 years. Partial spiral arms and arcs connecting the third and fourth winding show a larger variation in the wind velocity during these orbital periods on the timescale of ≈100 years.

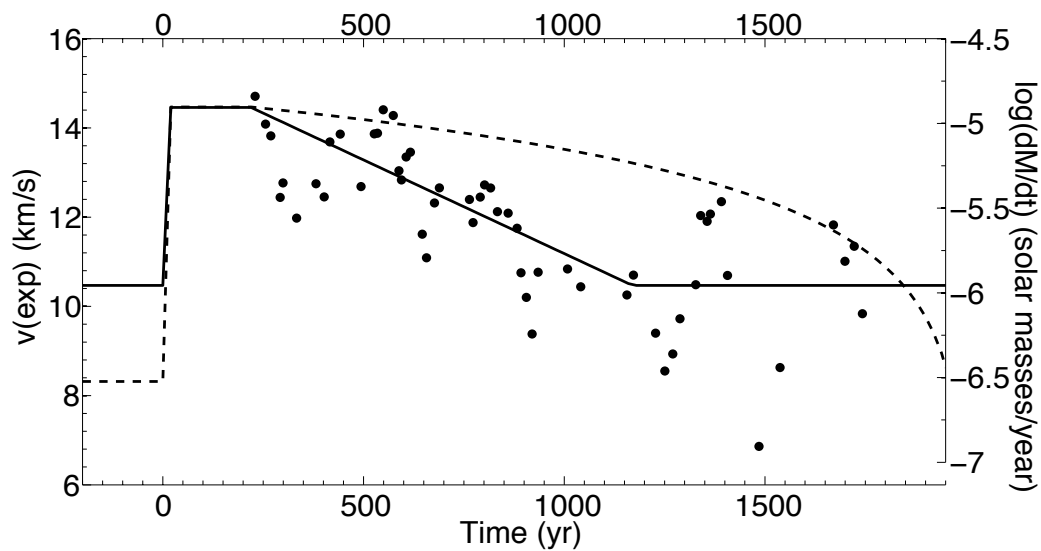

Figure 3: The velocity and mass-loss rate evolution of the stellar wind around R Sculptoris. The solid and dashed lines show the velocity and mass-loss rate as a function of time, respectively. The points correspond to the expansion velocities of the emission peaks in the p.a. vs. radius diagram (Fig. 2, left), assuming a binary

period of 350 years. The figure shows the evolution of the velocity and mass-loss rate since the onset of the last thermal pulse. The velocity profile is a fit to the data points, while the mass-loss-rate profile is constrained by the pre-pulse, pulse, and present-day mass-loss rates. The profiles are used as input to the SPH models. The shape of the mass-loss-rate profile is chosen to be consistent with the observations. The overall velocity fits the predictions from theoretical models of thermal pulses well. However, velocity variations of ±1.5 km s$^{-1}$ are apparent throughout the evolution, while theoretical models only predict significant variations in the expansion velocity <200 years after the pulse[4]. Theoretical models predict an increasing widening of the spiral with ± the sound speed, which may explain at least part of the velocity variations[8].

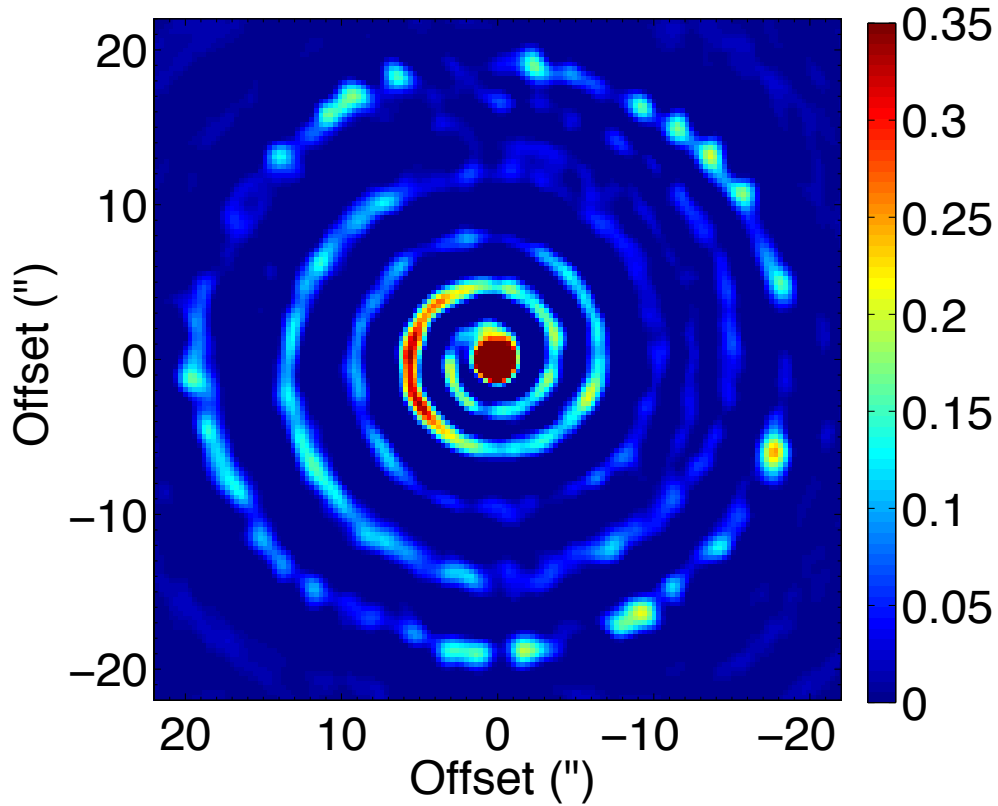

Figure 4: LIME radiative transfer model of the circumstellar structure around R Sculptoris. The model is based on the results from the SPH models at the stellar velocity. The results of the radiative transfer model have been processed by *simdata* in *CASA* using the ALMA Cycle 0 compact configuration specifications. The color scale is given in Jy/beam. The overall model intensities match the observed intensities well, while variations in the intensity contrast between the spiral windings and interwinding material indicate a more complicated mass-loss-rate variation. The inclination angle of the binary system to the line of sight is 90°. See supplementary information for an animation of the SPH models of R Sculptoris.